# Record-Breaking 1935.6 bit/s/Hz Spectral Efficiency in 19-Ring-Core Fiber Transmission of GMI-Estimated 25.24 Pb/s Capacity Using Low-Complexity 4×4 MIMO

Hualin Li[(1), †], Junyi Liu[(1), †], Jie Liu[(1), †, *], Shuqi Mo[(1)], Haolin Zhou[(1)], Yuming Huang[(1)], Yining Huang[(1)], Lei Shen[(2)], Shuo Xu[(2)], Lei Zhang[(2)], Jie Luo[(2)], Zhaohui Li[(1), *], Siyuan Yu[(1), *]

[(1)] School of Electronics and Information Engineering, State Key Laboratory of Optoelectronic Materials and Technologies, Sun Yat-sen University, Guangzhou 510275, China, liujie47@mail.sysu.edu.cn
[(2)] Yangtze Optical Fiber and Cable Joint Stock Limited Company, State key Laboratory of Optical Fiber and Cable Manufacture technology No.9 Guanggu Avenue, Wuhan, Hubei, P. R. China
†: These authors contributed equally.

**Abstract** *We achieve a record spectral efficiency of 1935.6 bit/s/Hz in the C+L bands in a 10-km 19-ring-core fiber supporting 266 OAM modes. GMI-estimated capacity of 25.24 Pb/s are transmitted using low-complexity 4×4 MIMO.* ©2025 The Author(s)

**Introduction**

Over the past four decades, single-mode fiber capacity has approached its nonlinear Shannon limit through innovations such as wavelength-division multiplexing and coherent detection [1]. Space-division multiplexing (SDM) has emerged as the next frontier, utilizing multi-core or few-mode fibers to increase space channels. Recent progress in SDM has demonstrated transmission rates exceeding 1-Pbps over dozens or even hundreds of spatial channels per fiber [2], making it particularly attractive for bandwidth-constrained scenarios like submarine cables and data center interconnects. However, densely packing spatial channels intensifies inter-channel crosstalk, necessitating large-scale MIMO digital signal processing with fast-growing complexity. As such, the spectral efficiency (SE) and the spectral efficiency density (SED) - defined as spectral efficiency per unit cross-sectional area of fiber – are both important figures-of-merit for evaluating the performance of SDM systems.

Fig. 1 illustrates the trade-off between the SE and SED and the MIMO processing complexity quantified by the required number of complex multiplications (RNCM) per bit [3]. The data are drawn from recently reported SDM/WDM transmission experiments where single-fiber capacities have surpassed 1 Pb/s. The figure reveals a clear trend: MIMO processing complexity increases with higher SED. This presents a fundamental challenge—achieving high spatial capacity density while keeping MIMO processing costs low is difficult in real-world communication systems.

To overcome this challenge, we present an SDM-WDM transmission system that achieves a record-breaking spectral efficiency (SE) of 1935.6 bit/s/Hz and a SED of 3.94×10$^{-2}$ bit/s/Hz/μm²(bidirectional transmission), requiring only 4×4 MIMO processing with time-domain equalization (TDE) tap count constrained within 35. Implemented over a 10-km 19-ring-core fiber (19-RCF) that supports 266 orbital angular momentum (OAM) mode channels derived from 19 cores × 7 OAM modes × 2 orthogonal polarizations, the transmission scheme employs probabilistically shaped 16QAM at a symbol rate of 12-GBaud in the C+L bands and QPSK in the S band, with wavelength channels spaced at 0.1 nm (12.5 GHz), ensuring efficient spectral utilization and high-capacity data transmission. To the best of our knowledge, this represents the first experimental demonstration of an SDM transmission system that simultaneously achieves high SE, high SED in a single fiber supporting more than 100 mode channels, while maintaining very low-complexity MIMO processing.

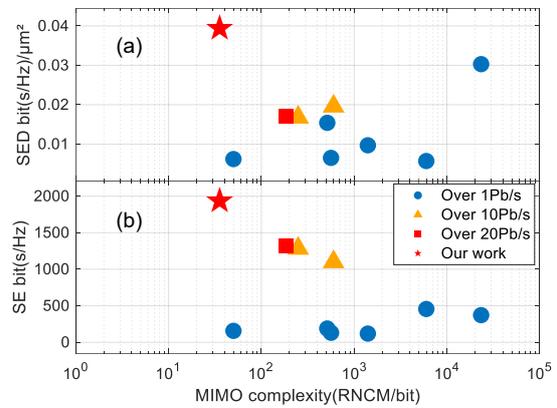

Fig.1. (a) SED of fiber in the C+L bands (GMI-estimated or before decoding) (b) SE in the C+L bands (GMI-estimated or before decoding) [2], [4-11].

**Experimental setup**

The experimental setup is shown in Fig. 2. At the transmitter, sliding test channels with a high optical signal-to-noise ratio (OSNR) and dummy channels with low OSNR are multiplexed to generate 1240 wavelength-division multiplexing (WDM) carriers. Three optical carriers, spaced at 0.1 nm / 12.5 GHz, from three tunable external cavity lasers (ECLs) serve as sliding test channels. The primary test wavelength is modulated by a dual-polarization IQ (DP-IQ) modulator, with a roll-off factor of 0.01, carrying a 12-Gbaud probabilistically shaped 16QAM signal in the C/L bands and QPSK in the S band. The primary channel is multiplexed with two additional wavelengths, which are modulated by another DP-IQ modulator with the same format. Both modulators are driven by a four-channel (AWG), generating signals of a $2^{18}-1$ pseudo-random binary sequence (PRBS). The dummy channels, derived from amplified spontaneous emission (ASE), is amplified by a thulium/erbium-doped fiber amplifier (T/E-DFA) and shaped by optical processors (OPs), setting the launch power of each WDM channel to 0 dBm.

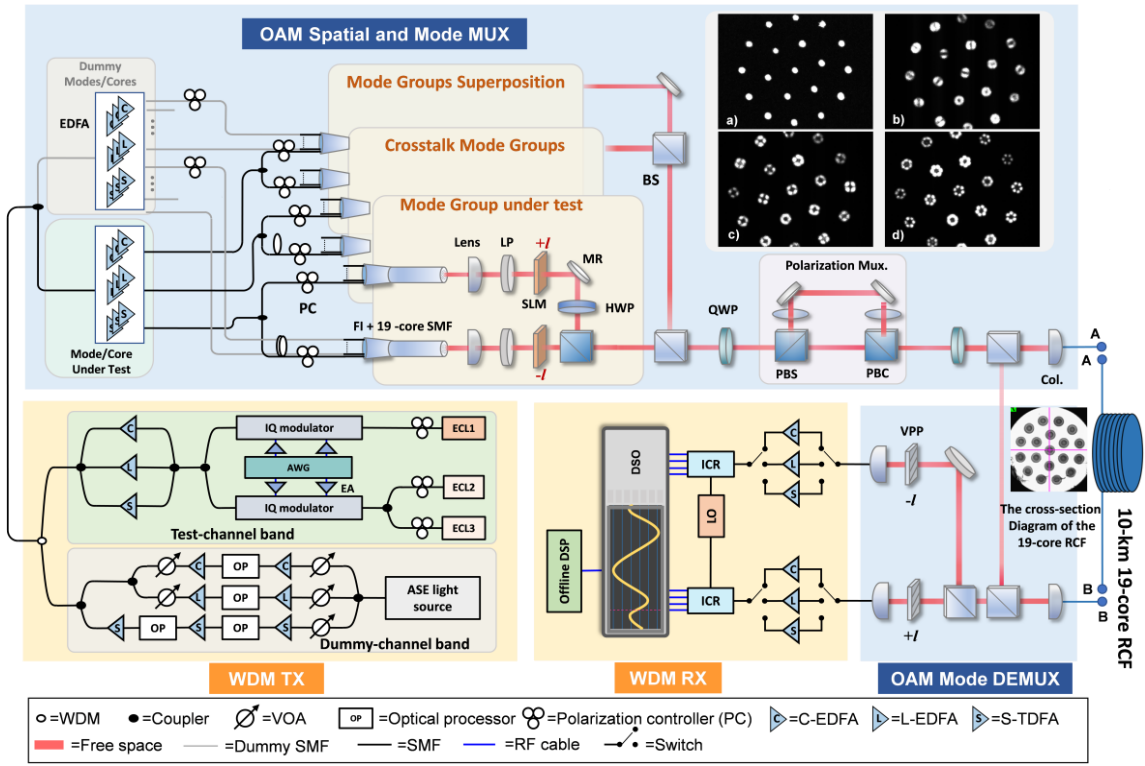

Fig.2. Experimental setup. ECL, external cavity laser; AWG, arbitrary waveform generator; EA, electrical amplifier; EDFA, erbium-doped fiber amplifier; LP, linear polarizer; SLM, spatial light modulator; MR, mirror; HWP, half-wave plate; BS, beam splitter; QWP, quarter-wave plate; PBS, polarization beam splitter; PBC, polarization beam combiner; Col., collimator; VPP, vortex phase plate; ICR, integrated coherent receiver. The intensity profiles of OAM MGs with $|l|$ = 0, 1, 2 and 3, which resemble linear polarization modes in each fiber core, arise from the coherent superposition of four orbital angular momentum modes within the mode group [12].

After multiplexing, the test and dummy channels are split into two paths, injected into the test cores and dummy cores of the fiber, respectively. The amplified signals are coupled into a 19-core SMF fan-in device, where the test core signals are amplified and decorrelated using optical fibers of varying lengths to ensure channel independence. The decorrelated signals are collimated and modulated by a spatial light modulator (SLM), generating orbital angular momentum (OAM) modes with topological charges $\pm l$ ($l$ = 0, 1, 2, 3). A four-focal length ($4f$) system combined with polarization beam splitters/combiners enables polarization multiplexing, producing four OAM modes: <+$l$, R>, <+$l$, L>, <−$l$, R>, and <−$l$, L> in each mode group (MG). The test core transmits three types of mode channels: test mode, adjacent crosstalk-inducing mode, and low-crosstalk background mode, forming a realistic crosstalk environment under limited resources. The dummy cores maintain the same excitation modes and power levels as the test core. Due to limited S-band amplifiers, identical power levels are applied only to the test core and adjacent cores, whereas the C+L band supports all cores. Prior studies confirm that inter-core crosstalk is negligible, ensuring minimal impact on communication signals [12].

The multiplexed OAM modes are coupled into a 10-km 19-RCF with a cladding diameter of 250 μm. Before fiber coupling, the beams are split into two branches via a 50:50 beam splitter (BS), injected from end A and end B, enabling bidirectional transmission. Prior experimental studies demonstrated that bidirectional transmission

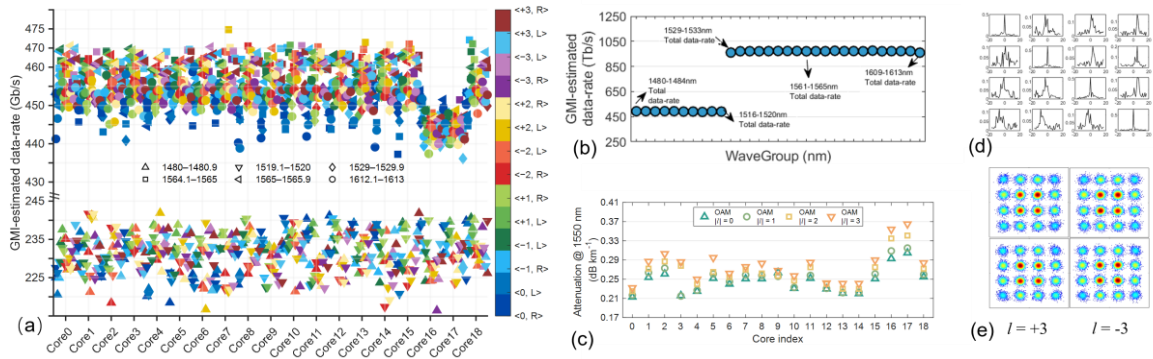

Fig3. (a) In the wavelength channels at the edges of the S, C, and L bands, the GMI-estimated data rate of 266 modes after transmission through 10 km of 19-RCF. (b)GMI-estimated data rata per 4nm range in the S, C and L bands. (c) The loss of the 19-RCF at 1550 nm. (d)MIMO tap weights of the topological charge |*l*| =3. (e) The constellations of PS-16QAM with |*l*| =3.

exhibits negligible Rayleigh backscattering effects [3]. After transmission, 19 petal-shaped OAM beam patterns are observed (see Fig. 2), attributed to weakly coupled excitation of inter-core and inter-mode components at the multiplexing stage. The emitted OAM beams are split into two paths by a BS, each passing through a vortex phase plate (VPP) with opposite topological charges, converted into Gaussian beams, then coupled into single-mode fibers (SMFs) and fed into dual-polarization integrated coherent receivers (ICRs).

At the transmitter, 329,840 channels (1240 wavelengths × 19 cores × 14 modes) are simultaneously multiplexed, but receiver hardware constraints limit demultiplexing to one mode group (MG) at a time, consisting of four OAM modes. These modes undergo 4 × 4 multiple-input multiple-output (MIMO) equalization to mitigate intra-MG crosstalk. The electrical waveforms from ICRs are sampled and stored by an eight-channel real-time oscilloscope (OSC) at 80 GSa/s. Offline DSP follows, including timing phase recovery, 4 × 4 MIMO equalization using the CMA, frequency offset estimation, and carrier phase estimation. Finally, GMI is calculated by soft decision.

**Results**

Fig. 3 presents a summary of data rate and transmission channel measurements. Fig. 3(a) illustrates the GMI-estimated data rate for all supported modes across the S, C, and L bands for each of the 19 fiber cores. The upper section represents the transmission capacity in the C+L bands, while the lower section corresponds to the S band, reflecting differences in modulation formats. Regardless of the transmission band, modes of the same order exhibit similar performance for each fiber core. Additionally, the figure highlights Core 16 and Core 17, which exhibit significantly lower data rates, attributed to higher mode loss (as shown in Fig. 3(c)). Since QPSK has a lower SNR requirement, the S-band transmission remains unaffected by the loss.

Fig. 3(b) summarizes the GMI-estimated data rate across the S, C, and L bands, evaluated at 4 nm intervals. Within each four-wavelength group, performance remains consistent, with C- and L-band channels supporting approximately 967.8 Tb/s, while S-band channels achieve around 491.8 Tb/s. The S-band limitation arises from instrumentation constraints, restricting its SNR to support only QPSK modulation. These results demonstrate high wavelength uniformity and core performance consistency, indicating that the transmission system maintains stable performance across different wavelengths and fiber cores.

**Conclusion**

We demonstrate an SDM-enabled ultra-wideband system spanning the S, C, and L bands achieving 25.24 Pb/s aggregate capacity (GMI-estimated). The bidirectional C+L bands transmission achieves a SE of 1935.6 bit/s/Hz with $3.94×10^{-2}$ bit/s/Hz/μm² SED realized with only 4×4 scale 35-tap TDE MIMO, representing the highest reported spatial efficiency and spatial efficiency density in a single-fiber system.

**Acknowledges**

This work was funded by National Key Research and Development Program of China (2024YFB2-908104); National Natural Science Foundation of China (62335019,62475294)

**References**

[1] Richardson, D. J., Fini, J. M., Nelson, L. E. "Space-division multiplexing in optical fibres," *Nature photonics* ,no.5,pp.354-362, 2013. DOI: 10.1038/NPHOTON.2013.94

[2] Liu, J., Zhang, J., Liu, J., Lin, Z., Li, Z., Lin, Z., Zhang, J., Huang, C., Mo, S., Shen, L. and Lin, S., "1-Pbps orbital

angular momentum fibre-optic transmission," *Light: Science & Applications*, *11*(1), p.202, 2022. DOI: 10.1038/s41377-022-00889-3

[3] Liu, J., Mo, S., Xu, Z., Huang, Y., Huang, Y., Li, Z., Guo, Y., Shen, L., Xu, S., Gao, R. and Du, C., "High spectral-efficiency, ultra-low MIMO SDM transmission over a field-deployed multi-core OAM fiber," *Photonics Research*, *13*(1), pp.18-30,2024.DOI: 10.1364/PRJ.533993

[4] Soma, D., Igarashi, K., Wakayama, Y., Takeshima, K., Kawaguchi, Y., Yoshikane, N., Tsuritani, T., Morita, I. and Suzuki, M., "2.05 Peta-bit/s super-nyquist-WDM SDM transmission using 9.8-km 6-mode 19-core fiber in full C band," In *2015 European Conference on Optical Communication (ECOC)* ,2015, pp. 1-3. DOI:10.1109/ECOC.2015.7341686

[5] Soma, D., Wakayama, Y., Beppu, S., Sumita, S., Tsuritani, T., Hayashi, T., Nagashima, T., Suzuki, M., Yoshida, M., Kasai, K. and Nakazawa, M., "10.16-Peta-B/s dense SDM/WDM transmission over 6-mode 19-core fiber across the C+ L band". *Journal of Lightwave Technology*, 36(6), pp.1362-1368,2018. DOI: 10.1109/JLT.2018.2799380

[6] Luís, R.S., Rademacher, G., Puttnam, B.J., Eriksson, T.A., Furukawa, H., Ross-Adams, A., Gross, S., Withford, M., Riesen, N., Sasaki, Y. and Saitoh, K., "1.2 Pb/s Throughput Transmission Using a 160 μm Cladding, 4-Core, 3-Mode Fiber," *Journal of Lightwave Technology*, *37*(8), pp.1798-1804.DOI:10.1109/JLT.2019.2902601

[7] Rademacher, G., Puttnam, B.J., Luís, R.S., Sakaguchi, J., Klaus, W., Eriksson, T.A., Awaji, Y., Hayashi, T., Nagashima, T., Nakanishi, T. and Taru, T., "10.66 peta-bit/s transmission over a 38-core-three-mode fiber," In *optical fiber communication conference*, pp. Th3H-1.

[8] Rademacher, G., Puttnam, B.J., Luís, R.S., Eriksson, T.A., Fontaine, N.K., Mazur, M., Chen, H., Ryf, R., Neilson, D.T., Sillard, P. and Achten, F., "Peta-bit-per-second optical communications system using a standard cladding diameter 15-mode fiber," *Nature Communications*, *12*(1), p.4238,2021.DOI: 10.1038/s41467-021-24409-w

[9] Rademacher, G., Luís, R.S., Puttnam, B.J., Fontaine, N.K., Mazur, M., Chen, H., Ryf, R., Neilson, D.T., Dahl, D., Carpenter, J. and Sillard, P., 1.53 peta-bit/s C-band transmission in a 55-mode fiber. In *2022 European Conference on Optical Communication (ECOC)* ,2022, pp. 1-4.

[10] Puttnam, B.J., Van Den Hout, M., Di Sciullo, G., Luis, R.S., Rademacher, G., Sakaguchi, J., Antonelli, C., Okonkwo, C. and Furukawa, H., "22.9 Pb/s data-rate by extreme space-wavelength multiplexing," In *IET Conference Proceedings CP839* ,2023, pp. 1678-1681. DOI: 10.1049/icp.2023.2665

[11] Rademacher, G., van den Hout, M., Luís, R.S., Puttnam, B.J., Di Sciullo, G., Hayashi, T., Inoue, A., Nagashima, T., Gross, S., Ross-Adams, A. and Withford, M.J., 2023, March. "Randomly coupled 19-core multi-core fiber with standard cladding diameter," In *Optical Fiber Communication Conference* ,2023,pp. Th4A-4.DOI: 10.1364/OFC.2023.Th4A.4

[12] Liu, J., Zhu, G., Zhang, J., Wen, Y., Wu, X., Zhang, Y., Chen, Y., Cai, X., Li, Z., Hu, Z. and Zhu, J., "Mode division multiplexing based on ring core optical fibers," *IEEE Journal of Quantum Electronics*, *54*(5), pp.1-18.2018. DOI:10.1109/JQE.2018.2864561